\input amstex.tex   \input amsppt.sty 

\pageheight{47pc}
\vcorrection{-1.1pc} 
\pagewidth{33pc}
\TagsOnRight 

\def\change#1\to#2\endchange{{\tt -change-} #1 {\tt -to-} #2 {\tt -end-}}
\def\add#1\endadd{{\tt -add-} #1 {\tt -end-}}

\magnification\magstep1

\PSAMSFonts

\def\RR{{\bold R}}
\def\CC{{\bold C}}
\def\SS{{\bold S}}
\def\HH{{\bold H}}

\def\Arg{\operatorname{Arg}}
\def\at{\char`@}
\def\Bu{B}
\def\gammat{{\widetilde\gamma}}
\def\Det{\operatorname{det}}
\def\e{\operatorname{e}}
\def\euc#1#2{\langle#1 \vert #2 \rangle_{{}_E}}
\def\FF{{\Cal F}}
\def\gh{{\widehat g}}
\def\lor#1#2{\langle #1 \vert #2 \rangle_{{}_L}}
\def\Lt{[L]}

\def\PP{{\Cal P}}
\def\scsss{simply connected symmetric symplectic space}
\def\sign{\operatorname{sign}}
\def\Sn{S}
\def\sss{symmetric symplectic space}
\def\St{{\widetilde S}}
\def\stress#1{{\it#1\/}}
\def\thetat{[\theta]}

\def\Yt{[Y]}

\topmatter

\author P. de M. Rios  \&  G.M.~Tuynman
\endauthor

\address PdMR: Laborat\'orio Nacional de Computa\c
c\~ao Cient\'\i fica; Av.~Get\'ulio Vargas 333; \hfill\break 
Petr\'opolis, RJ 25651-070; Brazil 
\endaddress

\curraddr Department of Mathematics; University of
California at Berkeley; \hfill\break 
Berkeley, CA 94720-3840; USA 
\endcurraddr

\email prios\at math.berkeley.edu \endemail

\address{GMT: CNRS UMR 8524 AGAT \&\ UFR de
Math\'ematiques;   Universit\'e de Lille I;\hfill\break 
F-59655 Villeneuve d'Ascq Cedex; France}\endaddress

\email Gijs.Tuynman\at univ-lille1.fr \endemail

\title On Weyl quantization from geometric quantization
\endtitle

\thanks This paper is a slightly  enlarged version of the
talk given at the XVII Bia\l owieza conference on
geometric methods in physics, Bia\l owieza (Poland),
1--8 July 1998.
\endthanks 

\abstract In \cite{We} a nice looking formula is
conjectured for a deformed product of functions on a
symplectic manifold in case it concerns a hermitian
symmetric space of non-compact type. We derive such a
formula 
for \scsss{}s
using ideas from geometric quantization and
prequantization of symplectic  groupoids. We compute the
result explicitly for the natural 2-dimensional
symplectic manifolds $\RR^2$, $\HH^2$, and
$\SS^2$. For $\RR^2$ we obtain the well known
Moyal-Weyl product. The other cases show that the
original idea in \cite{We} should be interpreted with
care. We conclude with comments on the status of our result.  

\endabstract

\endtopmatter

\refstyle{A}

\document 


\head Introduction \endhead

In \cite{We} the author discusses the quantization by
groupoids program as a means to obtain a
deformed multiplication of the Poisson algebra
$C^\infty(M)$ associated to a symplectic manifold $M$ in
the form $$
(fg)(z) = \int_{M \times M} f(x) g(y) K(x,y,z) \, dx\,dy
\ ,
$$
with a kernel $K_\hbar$, a function of a
deformation parameter $\hbar$, of the form $K_\hbar(x,y,z)
= \hbar^{-\dim M} \exp(iS(x,y,z)/\hbar)$, eventually
multiplied by an ``amplitude'' $A(x,y,z)$. 
It is suggested that for
hermitian symmetric spaces the function $S(x,y,z)$ should
be the symplectic surface of a geodesic triangle for
which the points $x$, $y$, and $z$ are the midpoints of
the sides.

In this paper we will derive such a formula (formula
\thetag5 below)  for \scsss{}s $M$
by means of geometric quantization of the
symplectic groupoid $M
\times M$ and its prequantization as described in
\cite{WX}. Our approach is inspired by the center-chord
representation on euclidean spaces as described in
\cite{OdA}. We then apply this procedure to three simple
2-dimensional  examples: the euclidean plane $\RR^2$, the
2-sphere $\SS^2\subset \RR^3$, and the hyperbolic plane
$\HH^2 \subset \RR^3$. The first example, already  worked
out in \cite{GBV} for $\RR^{2n}$, gives us the well known
Moyal-Weyl quantization of observables. In the hyperbolic
plane we see that we have to interpret the amplitude
function in a rather large sense: the phase function $S$ is defined only on
``half'' of $H^2 \times H^2 \times H^2$, forcing the
amplitude function to be zero outside this (open) domain,
and it blows up at the boundary of this domain. In the
example of the 2-sphere we encounter the additional
problem that midpoints do not always determine a unique
triangle.

\head The construction \endhead

\subhead Preliminaries \endsubhead

Let $(M,\omega)$ be a symplectic manifold and let $\hbar
\in \RR^+$ be a parameter. Let $(Y,\theta)$ be a
prequantization of $(M,\omega/\hbar)$, meaning that $\pi:Y
\to M$ is a principal $\SS^1$-bundle equipped with a
connection form $\theta$ whose curvature is $\omega/\hbar$
(which implies that the group of periods of $\omega$ is a
discrete subgroup of $\RR$). Using the identity
representation of the circle $\SS^1\subset \CC$ on $\CC$,
we let $L\to M$ be the associated complex line bundle
over $M$ with connection $\nabla$ and compatible
hermitian structure. It follows that we can identify $Y$
with the subset of $L$ of points of length 1 (with
respect to the hermitian structure). We now assume 
that the curvature of $\nabla$ also equals $\omega/\hbar$,
which implies that $\omega/\hbar$ represents an integral
cohomology class. This imposes a quantization condition on
$\hbar$ in case $\omega$ is not exact.

Our purpose is to construct a
map $C^\infty(M) \times C^\infty(M) \to C^\infty(M)$ by
means of geometric quantization of $M \times M$ as a
symplectic groupoid. Our strategy will be to use a
polarization such that the polarized sections of geometric
quantization can be identified with functions on $M$
(usually these sections form the Hilbert space, but here
we will interpret them as observables). Using a groupoid
structure on the prequantization, we construct the
looked-for product. To make this work, we will have to
restrict our attention to symplectic spaces with a 
complete affine connection
for which geodesic inversion with respect to a point
 is a symplectomorphism, and whose first homology group
is zero. This brings us in the category of \scsss{}s,
which includes all simply connected  hermitian
symmetric spaces. We will use extensively the results of
\cite{WX}, as well as its notation, but we will restrict
to the barest minimum of terminology. For more of that,
the interested reader is referred to \cite{WX} and the
references therein.

\subhead Prequantization of the groupoid \endsubhead

The construction starts by giving the manifold  $M
\times M$ the symplectic structure $(\omega, -\omega)$.
More precisely, if $\alpha$ and $\beta$ denote the
canonical projections $M \times M \to M$ onto the first and
second factor, then the symplectic form on $M \times M$ is
$\alpha^*\omega - \beta^*\omega$. The manifold $Y \times
Y$ is in a natural way a principal $\SS^1 \times
\SS^1$-bundle over $M \times M$. Quotienting out the
diagonal action of $\SS^1$ on $Y \times Y$,  $\e^{i\phi}
\cdot (y_1,y_2) = (\e^{i\phi} \cdot y_1,\e^{i\phi} \cdot
y_2)$, we obtain a principal $\SS^1$-bundle $\Yt = Y
\times Y / \SS^1 \to M \times M$. We will denote points in
$\Yt$ as $[y_1,y_2]$ with $y_i \in Y$. The induced
$\SS^1$-action is taken to be $\e^{i\phi}\cdot [y_1,y_2] =
[\e^{i\phi}\cdot y_1,y_2] = [y_1,\e^{-i\phi}\cdot y_2]$.
Moreover, the 1-form $(\theta,-\theta)$ induces a
connection form $\thetat \equiv [\theta,-\theta]$ on $\Yt$,
whose curvature is $(\omega/\hbar, -\omega/\hbar)$. We
thus obtain a (particular) prequantization of $M \times
M$. We let $\Lt \to M \times M$ be the associated complex
line bundle with connection and compatible
hermitian structure. And as before we identify $\Yt$ with
the subset of $\Lt$ of points of length 1. We define the
diagonal section $\varepsilon':M \to \Yt$ as
$\varepsilon'(m) = [y,y]$ with $y\in Y$ such that $\pi(y)
= m$. This section is horizontal for the connection
$\thetat$. It then follows from \cite{WX, theorem 3.1,
proposition 3.2} that there exists a unique groupoid
structure on $\Yt$ with given properties. In our case this
means that there exists a smooth map $\odot$ with values
in $\Yt$ and defined on pairs $[x,y_1], [y_2,z]\in \Yt$
such that $\pi(y_1) = \pi(y_2)$. Using that $\SS^1$ acts
transitively on the fibres of $Y\to M$, and the diagonal
$\SS^1$ action on $Y\times Y$, this condition means that
there exists a $z'\in Y$ such that $[y_2,z] = [y_1,z']$. 
With such a representation of the points, this
``multiplication'' $\odot$ is given by   
$$   
[x,y] \odot [y,z] = [x,z]  
\ . 
\tag1 
$$

\subhead The polarization \endsubhead

The next step in the geometric quantization procedure is
the choice of a polarization on $M \times M$. However,
for generic $M$ we know of no natural choices for a
polarization which ``mixes'' both factors $M$. We thus
try to find symplectic spaces for which we
can define a rather natural mixing polarization. Here is
the idea. For any complete affine connection $\nabla$ on
$M$ we can define a smooth map
$F : TM
\to M \times M$ by  
$$
F(m,v) = (\exp_m(-v), \exp_m(v))
\ ,
$$
where $\exp_m : T_mM \to M$ denotes the geodesic flow at
time $t=1$, starting at $m\in M$ and in the direction of
the tangent vector $v\in T_mM$. Since $\exp_m$ is a
diffeomorphism in a neighborhood of $0\in T_mM$, 
$F$ is a diffeomorphism in a neighborhood of the zero
section of $TM$. We define $U \subset TM$ as a maximal
connected and  symmetric (with respect to inversion in the
fibres of the tangent bundle) open neighborhood of the zero
section on which $F$ \stress{is} a diffeomorphism.
Associated to
$U$ we define $V = F(U) \subset M\times M$. Note that if
the (complete) affine connection $\nabla$ has no closed
geodesics, then $U=TM$ and $V= M\times M$. 

On $TM$ we have a
natural foliation $\FF_v$ whose leaves are just the fibres
$T_mM$ of the tangent bundle. Our idea is that its image
$\PP = F_*\FF_v$ should be a polarization for the
restriction of the symplectic form $(\omega, -\omega)$ to
$V$. An elementary computation shows that 
$\PP$ is a polarization on $V$ if and only if for
each $m\in M$ the map $\exp_m(v) \mapsto
\exp_m(-v)$ is a symplectomorphism on $\exp_m(U \cap T_mM)
\subset M$. We thus require that the symplectic manifold
$M$ admits a complete affine connection for which geodesic
inversion is a symplectomorphism. We thus arrive in the
category of \sss{}s \cite{Bi}\cite{RO}, which includes the category of
hermitian symmetric spaces because the connection
associated to the natural (complete) metric on a hermitian
symmetric space satisfies this condition.
When this condition is satisfied, we obtain a (real)
polarization
$\PP$ on $V\subset M\times M$. Moreover, as is obvious from
the definition of $\PP$ via $\FF_v$, the space of leaves
$V/\PP$ is naturally isomorphic to $M$, seen either as the
diagonal in $M\times M$ or as the zero section in $TM$.

We now claim that there exists a section $s_o : V \to
\Yt$ which is horizontal in the direction of $\PP$ and
which coincides with $\varepsilon'$ on its domain of
definition. The easiest way to construct this section is
by pulling back all structures on $M \times M$ to $TM$ by
means of the map $F$. More precisely, we define $\Omega$
as the closed 2-form $ F^*(\omega,-\omega)$ on $TM$ and
$(\Bu, \Theta)$ as  the principal $\SS^1$-bundle with
connection over $TM$ obtained by pulling back the bundle 
$(\Yt,\theta)$. Obviously the curvature form of $\Theta$
is $\Omega/\hbar$. As argued above, $\Omega$ is
identically zero on the fibres of $TM$, i.e., on the
leaves of $\FF_v$. The section $\varepsilon'$ of $\Yt$
gets transformed to a section $\sigma$ of $\Bu$ above $M$
seen as the zero section of $TM$. Since the fibres of
$TM$ are simply connected and since the curvature of
$\Theta$ is identically zero on these fibres, we can
extend the section $\sigma$ to a global section $TM \to
\Bu$ which is horizontal when restricted to a leaf of
$\FF_v$. Restricting this section to $U$ and then pushing
it to $V$ by means of $F$ we obtain our section as
claimed. 

In order to get a better grip on this section, let 
$(m_1,m_2) \in V \subset M \times M$ be arbitrary. We then
can define the curve $\gamma:[0,1] \to V \subset M \times
M$ by $\gamma(t) = F(m,tv)$, with $TM \supset U \ni (m,v) =
F^{-1}(m_1,m_2)$. More or less by construction
$s_o(m_1,m_2)$ is the end point of the horizontal lift of
$\gamma$ starting at $\varepsilon'(m)$. But the two
components $\gamma_1(t) =\exp_m(-tv)$ and $\gamma_2(t)
=\exp_m(tv)$ of the curve $\gamma$ form together the
geodesic from $m_1$ to $m_2$ with $m$ as midpoint.
Choosing $\mu\in \pi^{-1}(m)$ arbitrary, we thus can define
$\gammat_i(t)$ as the horizontal lift of $\gamma_i(t)$ in
$Y$ starting at $\mu$. Together they form a horizontal lift
in $Y$ above the geodesic between $m_1$ and $m_2$. By
definition of the connection form on $\Yt$, the curve
$\gammat(t) = [\gammat_1(t), \gammat_2(t)] \in \Yt$ is the
horizontal lift of $\gamma$ starting at $\varepsilon'(m)
= [\mu,\mu]$. It follows that $s_o(m_1,m_2) = [x,y]$ in
which $x$ and $y$ are the end points of a horizontal curve
above the geodesic (unique in $V$) between $m_1$ and
$m_2$. 

We continue with the geometric quantization program
and we look at the space of all sections of $\Lt$
above $V$ that are covariantly constant in the direction
of $\PP$. We will call such sections $\PP$-constant.
Viewing $\Yt$ as a subset of $\Lt$, the section $s_o$
constructed above is $\PP$-constant. Moreover, it
is a smooth nowhere vanishing section. It follows that
$\PP$-constant sections $s:V \to \Lt$ are in 1-1
correspondence with functions $f$ that are constant on
the leaves of $\PP$, i.e., with functions on $M = V/\PP$.
The identification is given by $s = f \cdot s_o$, or, more
precisely, by $s(m_1,m_2) = f(m_{12}) \cdot s_o(m_1,m_2)$,
where $m_{12}$ is the midpoint of the geodesic between
$m_1$ and $m_2$.

\subhead  A product of sections \endsubhead

We now stop the geometric quantization
program and we turn our attention to the groupoid
structure on $\Yt$. We extend the groupoid multiplication
$\odot$ to $\Lt$ by the following prescription. Any $p\in
\Lt$ can be written in a unique way as $p = \lambda
[x,y]$ with $\lambda \in [0,\infty)$ and $[x,y] \in \Yt
\subset \Lt$. Now, for $p_i =
\lambda_i [x_i,y_i]$ such that $\pi(y_1) = \pi(x_2)$ we
define  
$$
p_1 \odot p_2 =
\lambda_1 \lambda_2 [x_1,y_1]\odot[x_2,y_2]
\ .
$$ 
With this extended quasi-groupoid structure (quasi
because now not every element has an inverse), we 
construct a product on sections of $\Lt$. If
$s_1$ and
$s_2$ are two sections of $\Lt$ (not necessarily above
$V$, not necessarily $\PP$-constant), we
define a new section $s_1 \circledcirc s_2$ of $\Lt$ by
$$
(s_1 \circledcirc s_2)(m_1,m_3) = \int_M s_1(m_1,m_2)
\odot s_2(m_2,m_3) \,dm_2
\ .
$$
In this formula the measure $dm_2$ is the Liouville
measure on $M$ associated to the symplectic form
$\omega$. The integration makes sense because all
groupoid products $s_1(m_1,m_2)
\odot s_2(m_2,m_3)$ lie in the same fibre of $\Lt$~: the
one above $(m_1, m_3)$. Of course there is no guarantee
that this integral converges, but we will not deal with
these delicate analytical issues here. 

We now, for the moment, restrict our attention to the
case in which the metric $g$ has no closed geodesics,
i.e., the case in which $F$ is a diffeomorphism from $TM$
onto $M \times M$. In that case $\PP$-constant sections
of $\Lt$ are globally defined sections. For two
$\PP$-constant sections $s_i = f_i\cdot s_o$, $i=1,2$
with $f_i \in C^\infty(M)$ we thus get the formula
$$
(s_1\circledcirc s_2 )(m_1,m_3) = \int_M
f_1(m_{12}) f_2(m_{23}) s_o(m_1,m_2) \odot s_o(m_2,m_3)
\,dm_2
\ ,
\tag2
$$
in which $m_{jk}$ denotes the midpoint of the geodesic
between $m_j$ and $m_k$. Since $s_o$ is nowhere
vanishing, there must be a constant $\lambda$ such that
$s_o(m_1,m_2) \odot s_o(m_2,m_3) = \lambda
s_o(m_1,m_3)$. In order to determine this constant we
argue as follows. We  choose $x_1$, $x_2$, $x_3$, and
$x'_3$ such that $s_o(m_1,m_2)=[x_1,x_2]$,
$s_o(m_2,m_3)=[x_2,x_3]$, and $s_o(m_1,m_3)=[x_1,x'_3]$.
Note that we may take the same $x_1$ and $x_2$ beacuse of
the equivalence relation defining the points in $\Yt$. It
follows from formula \thetag1 that $s_o(m_1,m_2)
\odot s_o(m_2,m_3)$ equals $[x_1,x_3]$. But we know
that $x_1$ and $x_2$ are the endpoints of a horizontal lift
above the geodesic between $m_1$ and $m_2$, and similarly
for the pairs $x_2,x_3$ and $x_1,x'_3$. We thus have a
geodesic triangle $m_3m_2m_1$ and a horizontal lift
starting at $x_3$ above $m_3$, passing through $x_2$ and
$x_1$ and coming to $x'_3$, again above $m_3$. It follows
that $x'_3 = \lambda x_3$ with $\lambda\in \SS^1$ the
holonomy of the geodesic triangle $m_3m_2m_1$. In
particular we have $[x_1,x_3] = \lambda[x_1,x'_3]$. Now if
$\Delta(m_3m_2m_1)$ is any 2-chain whose boundary is the
geodesic triangle $m_3m_2m_1$, then $\lambda =
\exp(i\int_{\Delta(m_3m_2m_1)} \omega/\hbar)$. The result
does not depend upon the choice for $\Delta$ because the
curvature form $\omega/\hbar$ represents an integral
cohomology class. We are thus led to introduce the phase
function $\St(m_3,m_2,m_1) = \int_{\Delta(m_3m_2m_1)}
\omega$ representing the symplectic area of the 
surface $\Delta(m_3m_2m_1)$ whose boundary is the geodesic
triangle with corners at $m_3$, $m_2$, and $m_1$. Actually
$\St$ is in general multiple valued because there is (in
dimensions higher than 2) no unique such 2-chain $\Delta$,
but this indeterminacy disappears when taking the
exponential. On the other hand, in order to be sure that
such a 2-chain exists for all geodesic triangles, we
further restrict our attention to spaces $M$ without
homology in dimension 1. This
excludes for instance the 2-torus, but all simply
connected hermitian symmetric spaces satisfy this
condition, and thus in particular the hermitian symmetric
spaces of compact and non-compact type.

Thus, substituting these results in formula
\thetag2 we obtain 
${(f_1\cdot s_o
\circledcirc f_2 \cdot s_o)(m_1,m_3)} = $ \linebreak $g(m_1,m_3)\cdot
s_o(m_1,m_3)$, where $g$ is given by  
$$
g(m_1,m_3) = 
\int_M f_1(m_{12}) f_2(m_{23})
\e^{i\St(m_3,m_2,m_1)/\hbar}
\,dm_2 
\ .
\tag3
$$
If we forget the trivializing section $s_o$, we
thus have associated to two functions $f_1$, $f_2$ on $M$
a new function $g$ on $M \times M$.
In general, the product $s_1 \circledcirc s_2$ of two
$\PP$-constant sections will not be $\PP$-constant. In
terms of the function $g$ this means that, in general, the
function $g:M \times M \to \CC$ is not constant on the
leaves of $\PP$, i.e., of the form $g(m_1,m_3) =
\gh(m_{13})$ for some function $\gh:M \to \CC$  with
$m_{13}$ the midpoint of the geodesic between $m_1$ and
$m_3$.

\subhead A new product of functions on $M$ \endsubhead

In order to get a $\PP$-constant section, or, in other
words, in order to associate to two functions $f_1$ and
$f_2$ on $M$ a new function $f_1 \star f_2$ on $M$ (not on
$M\times M$), we integrate (average) over the leaves of $\PP$. This
is done most easily in terms of the fibres of $TM$ and we
get
$$
\aligned
(f_1 \star f_2)(m) &= 
\int_{T_{m}M} dv\ g(m_1,m_3) = 
\int_{T_{m}M} dv\ g(F(m,v))
\\ &=
\int_{T_{m}M} dv \int_M dm_2 \ 
f_1(m_{12}) f_2(m_{23})
\e^{i\St(m_3,m_2,m_1)/\hbar}
\ ,
\endaligned
\tag4
$$
with $(m_1,m_3) = F(m,v)$ and $m_{jk}$ the
midpoint on the geodesic between $m_j$ and $m_k$. It
remains to be decided what measure $dv$ to take on $T_mM$,
but there exists a rather canonical way to obtain one.
Using that $F$ is a global diffeomorphism (we are still
in that case), $F^*(\omega,-\omega)$ is a
symplectic form on $TM$, and thus we have its Liouville
volume form $d\mu_{TM}(m,v)$ on $TM$. On the other hand,
the zero section of $TM$ is diffeomorphic to the
symplectic manifold $(M, \omega)$, and thus on the zero
section of $TM$ we have its Liouville volume form
$d\mu_M(m)$. It follows that there exists a unique volume
form $dv_m(v) \equiv dv$ on each fibre $T_mM$ such that
$d\mu_M(m) \cdot dv_m(v) = d\mu_{TM}(m,v)$. 

Formula \thetag4 presents our deformed product of functions
on $M$. In order to write it in a nicer way,
we look at the map $\Psi : (v,m_2)
\mapsto (m_{12}, m_{23})$ from $T_{m}M \times M$ to
$M \times M$. We conjecture that this map is 
injective; it certainly need not be surjective as can be
seen in the case of the hyperbolic plane.  If we denote by
$dm_{12}$ the Liouville measure on the first factor of 
$M\times M$ and by $dm_{23}$ the Liouville measure on the
second factor, then there exists a positive function
$A_{m}$ on $W_{m} = \Psi(T_{m}M \times M)
\subset M\times M$ such that $\Psi^*(A_{m} \,
dm_{12}\, dm_{23}) = dv \, dm_2$. Associated to $W_m$ we
define the set $W\subset M^3$ as $W = \{ (m, m_{12},
m_{23}) \in M^3 \mid (m_{12}, m_{23}) \in
W_{m}\,\}$. We then can interpret the family of functions
$A_m$ as a single function $A: W \to [0, \infty)$ by
$A(m,m_{12}, m_{23}) = A_m(m_{12}, m_{23})$. 
Still under the assumption that $\Psi :
T_{m}M \times M \to W_{m}$ is bijective, we
define the function $\Sn$ on $W$ by $\Sn(m, m_{12},
m_{23}) = \St(m_3,m_2,m_1)$, where the points $m_i$ are
defined by the equations $(v,m_2) = \Psi^{-1}(m_{12},
m_{23})$ and
$F(m, v) = (m_1,m_3)$. The function $\Sn$ can be
described as the symplectic area of a surface $\Delta$
whose boundary is the geodesic triangle whose midpoints
of its three sides are $(m, m_{12},
m_{23})$. With these preparations, and denoting 
$m'\equiv m_{12}$ , $m''\equiv m_{23}$ , 
formula \thetag4 can be written as
$$
(f_1\!\star\! f_2)(m) \ = 
\iint_{W_{m}}
f_1(m') f_2(m'') \,
\e^{i\Sn(m,m',m'')/\hbar}
A({m},m', m'') \,
 dm'\, dm''
\ .
\tag5
$$
Except for the restriction of the integration to $W_m$
instead of $M\times M$, this is exactly of the form for a
deformed product as conjectured in \cite{We}.

\subhead The general case \endsubhead

We have derived formula \thetag5 under the assumption
that $F$ is a global diffeomorphism from $TM$ to $M\times
M$. If this is not the case, we were led to introduce the
subsets $U \subset TM$ and $V = F(U)\subset M\times M$, and
the section $s_o$ defined only above $V$. It follows that
the integration procedure which led us to formula \thetag3
can only be performed for those values of $m_2$ such that
$(m_1,m_2)$ and $(m_2,m_3)$ both lie in $V$. 
The next step of ``averaging'' over the leaves of $\PP$
should also be done with care. These
leaves are only defined in $V$ (elsewhere $\PP$ is not
defined), which means in terms of $TM$ that we
have to integrate, not over the whole tangent space
$T_mM$, but only over the part in
$U$, i.e., over $T_mM \cap U$. On the other hand, the
argument which led to the measure $dv$ remains valid: the
pull-back by $F$ of the Liouville measure on $V$ to $U$
gives us a measure on $U$. The zero section still carries
its natural Liouville measure, and thus there exists a
natural measure $dv_m$ on $T_mM \cap U$ such that it
completes the natural Liouville measure on the zero
section to the pull back of the Liouville measure on $V$.
We conclude that formula \thetag4 still defines a
deformed product of functions, provided we restrict
integration to the appropriate subset of $T_mM \times M$.

In the general case the map $\Psi$ need not be injective,
not even on the relevant subset $(T_mM \cap U) \times M$ as
described above, as can be seen in the example of the
2-sphere. However, inspired by the example of the
2-sphere, we conjecture that there still exists a
positive function $A_m$ on $W_m = \Psi((T_mM \cap U)
\times M)$ such that $\Psi^*(A_{m} \,
dm_{12}\, dm_{23}) = dv \, dm_2$. We also conjecture that
$\Psi$ is injective outside a closed subset of measure
zero in $(T_mM \cap U)\times M$. This means that we can
copy the arguments leading to formula \thetag5, and that
this formula is valid also in the general case, but with
the new subset $W_m$.

\pagebreak 

\head Three examples \endhead

\subhead The Euclidean plane $\RR^2$ \endsubhead

Let $M = \RR^2$ be the Euclidean plane with the
symplectic form $\omega = dp\wedge dq = d(p\,dq)$. The
(unique) prequantization is the bundle $Y = M \times
\SS^1$ with connection form $\hbar\theta = p\,dq +
d\varphi$. The map $F$, a global diffeomorphism, is given
as $F(p,q;v_p,v_q) = (p-v_p,q-v_q; p+v_p,q+v_q)$. A
horizontal lift of the curve $(p+tv_p,q+tv_q)$ is given
by $(q+tv_q, p+tv_p, exp(\tfrac i\hbar
(pt+\tfrac12t^2v_p)v_q))$. An elementary calculation then
gives for the section $s_o$ the expression 
$$
s_o(p_1,q_1; p_2,q_2) = [(p_1,q_1;1),(p_2,q_2;
\exp(\tfrac{i}{2\hbar}(p_1+p_2)(q_2-q_1))] 
\ ,
$$
where we used the equivalence relation on $[\ ,\ ]$ to
put the first phase equal to~1. From there it follows
immediately from formula \thetag1 that the phase factor
$\lambda$ in  $s_o(m_1,m_2) \odot s_o(m_2,m_3) = \lambda
s_o(m_1,m_3)$ is given by 
$$
\lambda = \exp(\tfrac i{2\hbar}\Bigl(
(p_1+p_2)(q_1-q_2) + (p_2+p_3)(q_2-q_3) + (p_3+p_1)(q_3-q_1)
\Bigr))
\ .
$$
A trivial calculation shows that this is indeed
$\exp(i\St(p_3,q_3; p_2,q_2;p_1,q_1)/\hbar)$ with $\St$ the
symplectic area (oriented with repect to the volume
form $dp\wedge dq$) of the triangle with corners at
$(p_3,q_3)$, $(p_2,q_2)$, and $(p_1,q_1)$. 

In this example, the change of coordinates $(v,m_2)
\mapsto (m_{12}, m_{23})$ is a linear bijection with
Jacobian $\frac14$, which implies that the amplitude
function
$A$ is constant~$\frac14$. Moreover, in the Euclidean
plane, the area  $\St(p_3,q_3; p_2,q_2;p_1,q_1)$ is four
times the area of the triangle determined by its
midpoints, i.e., $\Sn(p,q; p_{12},q_{12};p_{23},q_{23}) =
4 \St(p,q; p_{12},q_{12};p_{23},q_{23})$. Up to a scale
factor this result is the usual formula one gives for
Moyal-Weyl quantization of the Euclidean plane
(\cite{OdA}).

\subhead The hyperbolic plane $\HH^2$ \endsubhead

Our next example is the hyperbolic plane $\HH^2$ which we
interpret as one sheet of the 2-sheeted hyperboloid in
$\RR^3$ determined by the equations $z^2 - x^2 - y^2 = 1$
and $z>0$. We introduce the Lorentzian metric $\lor{\ }{\
}$ by the formula 
$$
\lor{(x,y,z)}{(x',y',z')} = zz' - xx' - yy'
\ .
$$
This metric induces a surface element, which we take as
symplectic form. 
An elementary but tedious calculation shows that the
oriented hyperbolic area of a triangle determined by its
three corners $a,b,c\in \HH^2 \subset \RR^3$ is given by
the formula 
$$
\St(a,b,c) = 2\Arg\Bigl(1+\lor ab + \lor bc + \lor ca +i
\Det(abc) \Bigr)
\ ,
\tag6
$$
where $\Arg$ denotes the argument of a complex number; it
lies in the interval $(-\pi,\pi)$. This formula is
derived in \cite{Ma} and \cite{Ur} in the context of
relativistic addition of velocities. 

The next steps are to express the area of a hyperbolic
triangle as a function of its midpoints and to determine
the change of coordinates $(v,m_2) \mapsto
(m_{12},m_{23})$. A straightforward
calculation shows that if $a,b,c\in \HH^2 \subset \RR^3$
are the corners of a hyperbolic triangle, and if
$\alpha,\beta,\gamma \in \HH^2
\subset \RR^3$ denote the midpoints of the three sides, 
then the area of the triangle (see \cite{Tu}, \cite{RO})
is given by the simple formula 
$$
\Sn(\alpha,\beta,\gamma) = 2\Arg\Bigl( \sqrt{1-
\Det(\alpha\beta\gamma)^2} + i \Det(\alpha\beta\gamma)
\Bigr)
= 2\arcsin(\Det(\alpha\beta\gamma))
\ .
\tag7
$$
The same analysis shows that the map $(a,b,c) \mapsto
(\alpha, \beta, \gamma)$ is injective onto the triples
$(\alpha, \beta, \gamma)$ satisfying  
$\Det(\alpha\beta\gamma)^2 < 1$, justifying the formula
for $\Sn$. It follows immediately that the subsets
$W_\alpha$ are given as 
$$
W_\alpha = \{ (\beta, \gamma)
\in \HH^2\times \HH^2 \mid \Det(\alpha\beta\gamma)^2 < 1
\, \}
\ . 
\tag8
$$
A final computation shows that the amplitude
function $A$ is given by 
$$
A(\alpha, \beta, \gamma) = 16  \lor\alpha\beta \cdot
\lor\beta\gamma \cdot \lor\gamma\alpha \cdot \Bigl( 1 -
\Det(\alpha\beta\gamma)^2 \Bigr)^{-5/2}
\ .
\tag9
$$
The fact that this amplitude function diverges on the
boundary of $W_\alpha$ shows that we correctly restricted
integration to this subset and that it is optimal.

\subhead The sphere $\SS^2$ \endsubhead

In the last example we consider the compact hermitian
symmetric space $\SS^2$ seen as the unit sphere in
$\RR^3$, i.e., determined by the equation $z^2 + x^2 +
y^2 = 1$. We equip $\RR^3$ with the Euclidean metric
$\euc{\ }{\ }$ given by
$$
\euc{(x,y,z)}{(x',y',z')} = zz' + xx' + yy'
\ .
$$
As for the hyperbolic plane, we take the induced surface
element as symplectic form. And again, an elementary but
tedious calculation shows that the oriented spherical area
of a triangle determined by its three corners $a,b,c\in
\SS^2 \subset \RR^3$ is given by the formula 
$$
\St(a,b,c) = 2\Arg\Bigl(1+\euc ab + \euc bc + \euc ca +i
\Det(abc) \Bigr)
\ ,
\tag10
$$
i.e., by exactly the same formula as in the hyperbolic
case, except that we use the Euclidean metric instead of
the Lorentzian one. However, this formula needs more
explanation than its hyperbolic counter part, because on
$\SS^2$ there are several triangles with the same three
corners. The area given by formula \thetag{10} is the area
of the triangle whose three corners are $a$, $b$, and $c$
and whose three sides all have length less than $\pi$. 

Elementary geometry shows that the subset $U\subset
T\SS^2$ is given by those tangent vectors that have
length less than $\pi/2$. In fact, if $v\in T_m\SS^2$ has
length $\pi/2$, the two points $\exp_m(-v)$ and
$\exp_m(v)$ are antipodal, and thus there is a circle of
pairs $(m,v)$ having these antipodal points as image under
$F$. It follows that the image $V = F(U)$ is the set of
pairs $(m_1,m_2)$ such that $m_1 \neq -m_2$. And indeed
for any two non-antipodal points there is a unique
geodesic with length less than $\pi$ joining them. 
The integration over $m_2$ in formula \thetag3 has to be
done over all those $m_2$ such that the two pairs $(m_1,
m_2)$ and $(m_2, m_3)$ belong to $V$. Since in the
definition of $V$ we only exclude antipodal points, this
means that we have to leave out a set of measure zero in
the integration over $m_2$. In other words, we can
maintain formula \thetag3 as it stands. The
factor $\e^{i\St(m_3,m_2,m_1)/\hbar}$ in the integration
over $m_2$ in \thetag3 is defined except on a set
of measure zero (when $m_2$ is antipodal to either $m_1$
or $m_3$). 

The integration over $v\in T_mM$ should not be done over
the whole of $T_mM$ but only over $T_mM \cap U$, i.e.,
over tangent vectors of length less than $\pi/2$.  That
this indeed corresponds exactly to integrating over the
leaves of $\PP$ can also be seen as follows. Two (pairs of)
points in  $V\subset
\SS^2 \times \SS^2$ lie on the same leaf of $\PP$ if and
only if they have the same midpoint on the geodesic
segment joining them. Since we avoid antipodal pairs,
there exists a unique geodesic segment of length less
than $\pi$ joining $(m_1,m_2)$, on which the midpoint is
given by the normalized average $(m_1+m_2) \cdot
(\euc{m_1+m_2}{m_1+m_2})^{-1/2}\in \SS^2$. We thus find
at the same time that the space of leaves is characterized
by $\SS^2$, the space of midpoints, and that the distance
of such a midpoint to one of its endpoints is less than
$\pi/2$, justifying the restriction to integrate only
over tangent vectors of length less than $\pi/2$.

It remains to express the phase function $\St$ in terms
of midpoints and to compute the amplitude function $A$.
Contrary to the hyperbolic case, there always exists a
geodesic triangle with given midpoints $\alpha, \beta,
\gamma \in \SS^2$. More precisely, if
$a,b,c\in \SS^2 \subset \RR^3$ are the corners of a
spherical triangle, and if $\alpha,\beta,\gamma \in \SS^2
\subset \RR^3$ denote the midpoints of the three sides,
then the oriented area $\Sn$ of the triangle is given as
(see \cite{Tu}, \cite{RO})
$$
\Sn(\alpha,\beta,\gamma) = 2\Arg\Bigl( \eta\sqrt{1-
\Det(\alpha\beta\gamma)^2} + i \Det(\alpha\beta\gamma)
\Bigr)
\ , 
\tag11
$$
where $\eta$ is a sign: the same as the majority of
signs among the three scalar products $\euc\alpha\beta$,
$\euc\beta\gamma$, and $\euc\gamma\alpha$ (provided they
are all non zero). We see that it is (up to the factor
$\eta$) the same formula as in the hyperbolic case.
Unlike the hyperbolic case, we do not have a restriction
on the midpoints, a fact which is corroborated by the
fact that for points on the unit sphere, the
determinant $\Det(\alpha\beta\gamma)^2$ is always less than
or equal to~1. 

However, the calculations leading to the formula for $\Sn$
show that, if all three sides of a triangle have length
less than $\pi$, then all three scalar products 
$\euc\alpha\beta$, $\euc\beta\gamma$, and
$\euc\gamma\alpha$ have the same sign, where the sign
should be interpreted as a function on $\RR$ defined as
being $+1$ for positive values, $-1$ for negative values,
and $0$ for zero. Thus, the set $W_\alpha$ is  
$$
W_\alpha = \{ (\beta,
\gamma) \in \SS^2 \times \SS^2 \mid 
\sign\euc\beta\gamma   = \sign
\euc\alpha\beta = \sign
\euc\alpha\gamma \, \}
\ . 
\tag12
$$
Moreover, the calculations also show that if all three
inner products are zero, then there is an infinity of
triangles having the given points as midpoints (roughly a
set parametrized by a point on $\SS^2$). But this set has
measure zero in $W_\alpha$ and hence can be
neglected in the integration. Note that even though the
triangle itself is not uniquely determined by its
midpoints, its area is.

Since $W_\alpha$ is only half of $\SS^2 \times
\SS^2$ (with respect to the natural measure), we 
have to take the restriction of the integration to
$W_\alpha$ in formula \thetag5 seriously. If the three
inner products do not all have the same sign, there still
exists a triangle $abc$ (unique if no inner product is
zero) but one of its sides will be longer than or equal
to $\pi$.

Computing
finally the amplitude function $A$ we find
$$
A(\alpha, \beta, \gamma) = 16\, \Bigl\vert
 \euc\alpha\beta \cdot
\euc\beta\gamma \cdot \euc\gamma\alpha 
\Bigr\vert
\cdot \Bigl( 1 -
\Det(\alpha\beta\gamma)^2 \Bigr)^{-5/2}
\ . 
\tag13 
$$

\head Conclusions \endhead 

Formula \thetag5 partly defines a deformed product of functions
on a symplectic manifold whose form was conjectured in
\cite{We}, in the spirit of the central (Weyl) representation 
of quantum observables \cite{OdA} and strict quantization \cite{Ri}.  
We have derived this formula using basic ideas from geometric 
quantization and groupoids, for
symplectic spaces without homology in dimension~1 and
which admit a complete affine connection for which geodesic
inversion is a symplectomorphism.  It 
should be noted that the final result does not
depend upon the choice of the prequantization bundle $Y$
for the symplectic manifold $M$. 

We emphasize the pure and simple geometrical character of the result 
(and its derivation). Accordingly, its algebraic  
and analytical  properties need 
to be clarified. Also, it should be compared with other approaches 
(e.g. \cite{Bi}\cite{BM}\cite{Ka}\cite{Urt}, see also \cite{Qi}). 
On the other hand, a striking property of formula \thetag5 is its asymptotic 
behaviour, specially when the functions are oscillatory: 
$f_1 \propto exp(ig_1 / \hbar)$ , $f_2 \propto exp(ig_2  / \hbar)$ . 
In this case, stationary phase evaluation of formula \thetag5 often yields  
$f_1\!\star\! f_2 \propto exp\{i(g_1\!\vartriangle\! g_2) / \hbar \}$ , 
where \linebreak
$(g_1\!\vartriangle\! g_2) (m)  = Stat_{(m',m'')} \{  
g_1 (m') + g_2 (m'') + S (m,m',m'')  \} $ 
is the composition of central generating functions (of canonical relations)  
$g_1$ and $g_2$ \cite{RO}. This is a rather promissing 
feature to be used in semiclassical analysis.

\head Acknowledgements \endhead

We thank S.~Berceanu, M.~Bordemann, A.M.~Ozorio~de~Almeida and A.~Weinstein 
for stimulating discussions and helpful remarks.
We also acknowledge support from CNPq.

\pagebreak


\widestnumber\key{GBV}
\Refs

\ref
\key Bi 
\by P. Bieliavsky 
\paper Strict quantization of solvable symmetric spaces 
\jour arXiv:math.QA/0010004 
\endref   

\ref
\key BM 
\by P. Bieliavsky \& M. Massar 
\paper Strict deformation quantization for actions of a class of symplectic lie groups   
\jour arXiv:math.QA/0011144  
\endref   

\ref
\key GBV
\by J.M. Gracia-Bond\'\i a \& J.C. V\'arilly
\paper From geometric quantization to Moyal quantization
\jour J. Math. Phys.
\vol 36
\yr1995
\pages2691--2701
\endref

\ref
\key Ka 
\by M.V. Karasev  
\paper Geometric Star Products 
\jour Cont. Math.
\vol179
\yr1994
\pages 115--121
\endref

\ref
\key Ma
\by A.J.~Macfarlane
\paper On the restricted Lorentz group and groups
homomorphically related to it
\jour J. Math. Phys.
\vol 3
\yr1962
\pages1116--1129
\endref


\ref
\key OdA
\by A.M. Ozorio de Almeida
\paper The Weyl Representation in Classical and Quantum
Mechanics 
\jour Phys. Rep.
\vol 295
\yr1998
\endref

\ref 
\key Qi 
\by Z. Qian 
\paper Groupoids, midpoints and quantization 
\jour PhD thesis, U.C.Berkeley 
\yr 1997 
\endref 

\ref 
\key Ri 
\by M.A. Rieffel 
\paper Deformation quantization of Heisenberg manifolds 
\jour Commun. Math. Phys. 
\vol122
\yr1989
\pages531--562 
\endref


\ref
\key RO
\by P.~de~M.~Rios \& A.~Ozorio de Almeida
\paper A Variational Principle for Actions on Symmetric
Symplectic Spaces
\miscnote Preprint. Revised and edited english translation of PhD thesis - CBPF, Rio de Janeiro, 2000 
\endref

\ref
\key Tu
\by G.M.~Tuynman
\paper Areas of spherical and hyperbolic triangles in
terms of their midpoints
\yr1999
\miscnote preprint
\endref

\ref
\key Ur
\by H.~Urbantke
\paper Physical holonomy, Thomas precession, and Clifford
algebra
\jour Am. J. Phys.
\vol 58
\yr 1990
\pages747--750
\moreref
\paper Erratum
\jour Am. J. Phys.
\vol59
\yr1991
\pages1150--1151
\endref 

\ref
\key Urt
\by A. Urtenberger 
\paper Quantization, symmetries and relativity 
\jour Cont. Math. 
\vol 214 
\yr 1996 
\pages 169--187 
\endref 


\ref
\key We
\by A. Weinstein
\paper Traces and Triangles in Symmetric Symplectic Spaces
\jour Cont. Math.
\vol179
\yr1994
\pages261--270
\endref


\ref
\key WX
\by A. Weinstein \& P. Xu
\paper Extensions of symplectic groupoids and quantization
\jour J. reine angew. Math.
\vol417
\yr1991
\pages159--189
\endref


\endRefs
\enddocument